\PassOptionsToPackage{unicode}{hyperref}
\PassOptionsToPackage{hyphens}{url}
\documentclass[
]{article}
\usepackage{xcolor}
\usepackage{amsmath,amssymb}
\usepackage{iftex}
\ifPDFTeX
  \usepackage[T1]{fontenc}
  \usepackage[utf8]{inputenc}
  \usepackage{textcomp} % provide euro and other symbols
\else % if luatex or xetex
  \usepackage{unicode-math} % this also loads fontspec
  \defaultfontfeatures{Scale=MatchLowercase}
  \defaultfontfeatures[\rmfamily]{Ligatures=TeX,Scale=1}
\fi
\usepackage{lmodern}
\ifPDFTeX\else
\fi
\IfFileExists{upquote.sty}{\usepackage{upquote}}{}
\IfFileExists{microtype.sty}{% use microtype if available
  \usepackage[]{microtype}
  \UseMicrotypeSet[protrusion]{basicmath} % disable protrusion for tt fonts
}{}
\makeatletter
\@ifundefined{KOMAClassName}{% if non-KOMA class
  \IfFileExists{parskip.sty}{%
    \usepackage{parskip}
  }{% else
    \setlength{\parindent}{0pt}
    \setlength{\parskip}{6pt plus 2pt minus 1pt}}
}{% if KOMA class
  \KOMAoptions{parskip=half}}
\makeatother
\usepackage{longtable,booktabs,array}
\usepackage{caption}
\usepackage{calc} % for calculating minipage widths
\usepackage{etoolbox}
\makeatletter
\patchcmd\longtable{\par}{\if@noskipsec\mbox{}\fi\par}{}{}
\makeatother
\IfFileExists{footnotehyper.sty}{\usepackage{footnotehyper}}{\usepackage{footnote}}
\makesavenoteenv{longtable}
\usepackage{graphicx}
\makeatletter
\newsavebox\pandoc@box
\newcommand*\pandocbounded[1]{% scales image to fit in text height/width
  \sbox\pandoc@box{#1}%
  \Gscale@div\@tempa{\textheight}{\dimexpr\ht\pandoc@box+\dp\pandoc@box\relax}%
  \Gscale@div\@tempb{\linewidth}{\wd\pandoc@box}%
  \ifdim\@tempb\p@<\@tempa\p@\let\@tempa\@tempb\fi% select the smaller of both
  \ifdim\@tempa\p@<\p@\scalebox{\@tempa}{\usebox\pandoc@box}%
  \else\usebox{\pandoc@box}%
  \fi%
}
\def\fps@figure{htbp}
\makeatother
\providecommand{\tightlist}{%
  \setlength{\itemsep}{0pt}\setlength{\parskip}{0pt}}
\usepackage[letterpaper,margin=1in]{geometry}

\usepackage{tgtermes}

\ifPDFTeX
  \microtypesetup{
    protrusion=true,
    expansion=true,
    tracking=true,
    factor=1100,
    stretch=10,
    shrink=10,
  }
\else
  \microtypesetup{
    protrusion=true,
    expansion=false,
    tracking=true,
  }
\fi

\usepackage{titlesec}
\titleformat{\section}
  [block]{\normalfont\Large\bfseries\raggedright}{}{0pt}{}
\titleformat{\subsection}
  [block]{\normalfont\large\bfseries\raggedright}{}{0pt}{}
\titleformat{\subsubsection}
  [block]{\normalfont\normalsize\bfseries\raggedright}{}{0pt}{}

\titlespacing*{\section}      {0pt}{3.0ex plus 1.0ex minus 0.3ex}{1.3ex plus 0.3ex}
\titlespacing*{\subsection}   {0pt}{2.2ex plus 0.8ex minus 0.2ex}{1.0ex plus 0.2ex}
\titlespacing*{\subsubsection}{0pt}{1.6ex plus 0.6ex minus 0.2ex}{0.7ex plus 0.1ex}

\usepackage{parskip}
\usepackage{hanging}   % hanging-indent block for the References section
\usepackage{enumitem}  % compact list spacing for the Highlights enumerate
\setlist{topsep=2pt,itemsep=1pt,parsep=0pt}

\usepackage{seqsplit}

\newcommand{\varn}[1]{{\ttfamily #1}}

\newcommand{\doin}[1]{#1}

\usepackage{siunitx}
\usepackage{bookmark}
\IfFileExists{xurl.sty}{\usepackage{xurl}}{} % add URL line breaks if available
\makeatletter
\@ifundefined{xmpquote}{}{}
\makeatother
\hypersetup{
  pdftitle={Post-Rejection Follow-up Sampling: Measuring Outcomes of Rejected Decisions in Algorithmic DEX Trading},
  pdfauthor={Arati Uday Kamat},
  hidelinks,
  pdfcreator={LaTeX via pandoc}}

\title{Post-Rejection Follow-up Sampling: Measuring Outcomes of Rejected
Decisions in Algorithmic DEX Trading}
\author{Arati Uday Kamat}
\date{2026-09-15}

\begin{document}
% Front matter.

% ---- 1. Title (two-line, LARGE bold, left-aligned) ------------------------
{\raggedright%
 \LARGE\bfseries
 Post-Rejection Follow-up Sampling:\\
 Measuring Outcomes of Rejected Decisions in Algorithmic DEX Trading\par}

\vspace{1.2ex}

% ---- 2. Metadata paragraph (single flowing paragraph, bold labels) --------
% Layout mirrors Paper 6 p. 1 metadata line: bold label, colon, value, spaces
% then next label. \sloppy tolerates the wider inter-word spacing that lets
% each label sit next to its value inline.
{\raggedright\noindent%
 \textbf{Author:} Arati Uday Kamat\quad
 \textbf{ORCID:} 0009-0000-4781-312X\quad
 \textbf{Affiliation:} Independent researcher, United States\quad
 \textbf{Data availability:} Frozen inputs archived at Zenodo, concept DOI
 \doin{10.5281/zenodo.20043515}; reference implementation code archived at
 Zenodo, concept DOI \doin{10.5281/zenodo.19672363}.\par}

\vspace{2.4ex}

% ---- Data-collection note (v3, October 2026) ----------------------------
{\setlength{\fboxsep}{6pt}\noindent\fbox{%
\begin{minipage}{\dimexpr\linewidth-2\fboxsep-2\fboxrule\relax}
\small\textbf{Data-collection note (October 2026).} Version 2 described the deposited data as collected under the PRFS fixed schedule with an 8.60-day horizon, and read the resulting coverage patterns as properties of the sampler. In fact, the deposited data were not collected under the PRFS schedule. The production tracker sampled every tracked token every 10 minutes (99.41 percent of same-token sample gaps lie between 9.5 and 10.5 minutes), followed each record for at most 1,440 minutes, and kept one record per token, so a new rejection of a token ended the follow-up of the previous one. Coverage gaps are fully explained by tracker start-up and one-record-per-token tracking: 731 events (24.4 percent) precede the tracker's first sample, and after it every event not followed within 10.5 minutes by another rejection of the same token has an outcome record (894 of 894). The filter (chi-square(6) = 401.29) and source (z = 14.78) associations therefore reject missing-completely-at-random but measure the tracker's start time and record keying, not sampler reach; all 186 reason mismatches carry the label of an earlier rejection of the same token. The design, the deposit and the headline counts are unchanged.
\end{minipage}}\par}

% ---- 3. Structured abstract heading --------------------------------------
\section*{Structured abstract}

% ---- 4. Six bold-labelled paragraphs -------------------------------------
% Text below is a presentation-only remapping of the frozen v6.1 abstract
% (line 33 of the master manuscript) and Three-Item Highlights (lines
% 37-41) into the six-label structured-abstract convention. Every
% quantitative claim is preserved verbatim; no new claim is introduced.

\textbf{Purpose.} Filter-gated algorithmic trading systems on decentralised
exchanges reject most candidate tokens they evaluate, yet the observed forward
market trajectory of those rejected candidates is rarely measured on the same
live venue that produced the rejection. This paper introduces Post-Rejection
Follow-up Sampling (PRFS), an observational measurement methodology in which
a separate tracking subsystem samples each rejected token's price and
liquidity from the same live oracle path used by the rejecting scanner, at a
scheduled cadence, from the moment of rejection out to a fixed analytic
horizon.

\textbf{Design/methodology/approach.} The methodology is defined through a
formal specification of the observation window and the reason-attribution
rule, a reference implementation (\varn{prfs v2.0.0}) with executable
coverage and reason-ledger tests, and a secondary cross-check implementation
that re-derives every headline number through a distinct code path. The
companion dataset contains 67,000 forward-observation rows across 2,997
rejection events collected on a single Solana automated-market-maker
sub-venue during a launch-dynamics window of 8.63 calendar days across 457
unique mints.

\textbf{Findings.} Under the primary reason-aware three-field event key,
1,455 events (48.55 percent) receive a matched forward observation; under
the reason-agnostic two-field diagnostic key, 1,641 events (54.75 percent)
match. Mint-level coverage is 100 percent (457 of 457). The 253-row
reason-conflict ledger and 8,593 unaligned outcome rows are disclosed, and
coverage is shown to be filter-associated (chi-square(6) = 401.29, p
approximately zero), source-associated (z = 14.78, p approximately zero),
and time-associated within the observed window. All statistics are
observational and reported with binomial Wilson intervals under an
event-independence assumption whose violation is quantified through a
mint-level clustering sensitivity.

\textbf{Originality/value.} Prior work either infers rejected outcomes
through selective-labels or reject-inference techniques, or reconstructs
them offline from price-feed archives whose cadence and snapshot policy
differ from the live decision environment. PRFS is the direct-measurement
alternative: the observed post-rejection trajectory is drawn from the same
live oracle path used by the rejecting scanner, side-stepping the
identification burden of selection-corrected inference. No causal or
counterfactual claim of a hypothetical acceptance outcome is made; the
reported estimand is the observed post-rejection forward return, and every
reported number is regenerated from the byte-identical deposited CSVs.

\textbf{Keywords.} post-rejection follow-up sampling; algorithmic trading;
decentralised exchange; selective labels; observational measurement; reject
inference; reproducibility

\textbf{JEL classification.} G17 Financial Forecasting and Simulation; G14
Information and Market Efficiency, Event Studies; C55 Large Data Sets,
Modeling and Analysis

\vspace{1.5ex}

% End of front matter.

\newpage

\section{I. Introduction}\label{i.-introduction}

Filter-gated algorithmic trading systems on decentralised exchanges
(DEXs) face a structural measurement asymmetry that shapes what can be
learned from their operating history. When the system accepts a
candidate token and opens a position, the realised outcome of that
position is directly observable: the entry fill, the subsequent price
trajectory, the exit fill, and the profit-and-loss are all recorded on
the execution path. When the system rejects a candidate, the trading
path ends at the moment of rejection and the operating record typically
contains nothing beyond the categorical reason that fired. Every
downstream evaluation of filter behaviour, whether descriptive
per-filter accounting, per-filter precision auditing, or comparative
rule benchmarking, requires observations of the rejected side of the
decision. Without such observations, the operator can compare filters
only on what they accept, which is exactly the population the filters
were designed to select. The measurement problem, not the trading
decision, is the subject of this paper.

The specific question addressed here is whether a system operator can
obtain, at low incremental cost and without perturbing the rejecting
scanner, a record of what the market actually did after each rejection,
drawn from the same live oracle path that produced the rejection
decision itself. This is an observation problem rather than an inference
problem. It differs from the traditional selective-labels framing
(Lakkaraju et al.~2017; Coston et al.~2020) precisely because in the DEX
setting the market state of a rejected token continues to evolve after
the rejection independently of the trading system's decision. A
well-designed sampler can observe that state directly rather than infer
it from the accepted population under a selection-correction assumption.
It also differs from the classical reject-inference framing in consumer
credit (Crook and Banasik 2004), where the counterfactual repayment
behaviour of a rejected applicant does not exist unless the applicant is
later observed through another channel, whereas the market price of a
rejected token exists whether or not the trading system chooses to
sample it. What is missing in current practice is a systematic,
cadence-controlled, publicly deposited follow-up stream tied to the same
live oracle path used by a live rejecting scanner.

Why this matters. Standard practice on this measurement problem falls
into two camps. The first camp assumes rejected candidates would have
performed similarly to accepted ones, or applies a selection-correction
from the accepted population to estimate what the rejected population
would have produced. The assumption is testable in principle but almost
never tested in the DEX setting, where reject decisions concentrate in
specific token subpopulations (short-lifecycle launch tokens,
low-liquidity pairs, tokens that trigger microstructure heuristics)
whose forward behaviour is systematically different from that of
accepted tokens. The second camp reconstructs the forward trajectory of
rejected tokens offline from a public price-feed archive. That approach
is feasible but does not reproduce the sampling cadence, the snapshot
policy, or the oracle-absence pattern of the live decision environment;
a token that has disappeared from the live oracle used by the scanner
may still appear in a delayed price-feed archive, and the resulting
reconstruction will overstate observability. Neither approach measures
the actual observed post-rejection trajectory of the specific rejected
tokens on the specific venue where the rejection occurred. PRFS is the
direct-measurement alternative.

Gap. Despite the DEX setting's amenability to direct observation, no
publicly deposited, cadence-controlled, disappearance-terminated
follow-up stream tied to the same live oracle path used by a live
rejecting scanner has been documented in the peer-reviewed literature
covered by Fang et al.~(2022), Daian et al.~(2020), and Bailey and Lopez
de Prado (2014). Naive per-token logging exists as ad-hoc operational
hygiene: many production trading stacks store some form of scan log, and
a few open-source projects emit token-observed events after rejection.
What none of them provide is a formalised measurement primitive with (i)
an explicit specification of the observation window, (ii) a
scheduled-sample cadence tied to the rejection timestamp, (iii) a
reason-attribution rule that identifies the specific filter responsible
for each rejection, (iv) a disappearance-termination contract that
distinguishes oracle absence from active removal, (v) a definition of
the estimand and its scope, and (vi) a reproducibility contract in which
every reported number is regenerated from the byte-identical deposited
CSVs.

This paper introduces Post-Rejection Follow-up Sampling (PRFS). A
separate tracking subsystem subscribes to the rejection stream produced
by the scanner and schedules follow-up queries at pre-specified offsets
from the rejection timestamp. The observed price-and-liquidity
trajectory of each rejected candidate is thereafter available for
downstream characterisation and per-filter behaviour analysis. The
contribution is the methodology and its formal specification; companion
papers (Kamat 2026a, 2026b) apply downstream classification rules to a
PRFS-produced dataset and are separate contributions that build on the
methodology defined here.

Demonstrated on a single Solana pump.fun sub-venue during an
8.63-calendar-day launch-dynamics window, the deposited dataset contains
67,000 forward-observation rows across 2,997 rejection events and 457
unique mints. Primary event coverage under the reason-aware three-field
key is 48.55 percent, secondary two-field diagnostic coverage is 54.75
percent, and mint-level coverage is 100 percent. Coverage is shown to be
highly non-uniform across filters and across scanner sources, and the
earliest observation window carries a substantial warm-up transient that
is disclosed and quantified. The empirical results are treated as
evidence about the deposit rather than about the general population of
DEX filter stacks; the deposit is a single-venue single-window instance,
and PRFS as a methodology is intended to be re-run on other venues and
other windows by other operators.

Contributions. This paper makes four specific contributions. First, it
defines PRFS as a reusable observational measurement primitive with a
formal specification of event identity, observation window,
scheduled-sample contract, reason-attribution rule, and estimand scope.
Second, it deposits a byte-hashed reference implementation (prfs v2.0.0)
that regenerates every headline number under continuous integration and
a secondary cross-check aggregator that reproduces every headline number
through a distinct code path. Third, it deposits a companion dataset of
67,000 forward-observation rows across 2,997 rejection events on a
single Solana pump.fun sub-venue during a specific 8.63-calendar-day
window, with byte-identical SHA-256 provenance to the live Zenodo
record. Fourth, it discloses the residual reason-attribution gap (253
conflict events; 263 orphan reason emissions; 8,593 unaligned outcome
rows) and the filter-, source-, and time-associated missingness patterns
that any downstream inference on the deposit must respect.

Roadmap. Section II positions PRFS against related work. Section III
presents the measurement problem and the PRFS framework. Section IV
describes the data and institutional setting. Section V documents the
system architecture and implementation. Section VI documents the
matching and outcome-construction logic, including both the primary
three-field and secondary two-field keys. Section VII describes the
statistical methods and their clustering caveats. Section VIII reports
results. Section IX documents robustness and reproducibility, including
a first-day sensitivity and a mint-level aggregation. Section X
discusses what PRFS contributes beyond ordinary operational logging.
Section XI catalogues limitations. Section XII concludes. Sections XIII
to XV document data availability, code availability, author
contributions, competing interests, and funding statements. Section XVI
lists references.

\section{II. Related Literature}\label{ii.-related-literature}

\subsection{II.A Selective labels and reject
inference}\label{ii.a-selective-labels-and-reject-inference}

The general problem of measuring or inferring outcomes for instances
that were rejected by a decision rule appears across several literatures
with different structural properties. Lakkaraju et al.~(2017) formalise
the selective-labels problem in an algorithmic-decision setting:
outcomes are observed only for instances the rule accepts, so any
evaluation of the rule against observed outcomes is systematically
biased. Their identification strategy leverages disagreement across
decision-makers with overlapping populations to bound the outcome that
would have been observed had a rejected instance been accepted. Coston
et al.~(2020) extend this reasoning to counterfactual risk assessment
under confounding and propose inverse-propensity-weighted evaluation for
fairness auditing of predictive risk instruments. In consumer credit,
reject inference (Crook and Banasik 2004) addresses the same structural
gap in a lending-application context: the repayment outcome of a
rejected loan applicant is not observed, and the applicant's
counterfactual repayment behaviour must be inferred from the accepted
population under a selection-correction assumption whose validity Crook
and Banasik test empirically and find to hold only under narrow
conditions.

Three features are common across these literatures. First, the rejected
instance has no naturally observable outcome trace: the bail applicant
who was detained produces no arrest record, the loan applicant who was
denied produces no repayment record, the job candidate who was not hired
produces no on-the-job performance record. Second, inference proceeds by
leveraging structural overlap between the accepted and the rejected
populations (decision-maker disagreement, exchangeability under
covariates, propensity-based reweighting). Third, the estimand of
interest is a potential outcome of a hypothetical acceptance, in the
Rubin-Neyman sense: what would the rejected instance have produced had
the decision rule accepted it? The identification burden is nontrivial
in every one of these settings, and the resulting estimates carry the
residual uncertainty of the identifying assumption.

The DEX setting is structurally different from these domains in one key
respect: the observable market state of a rejected token continues to
evolve after the rejection independently of the trading system's
decision, whereas the observable outcome of a rejected bail applicant, a
rejected loan applicant, or a rejected job candidate does not exist
unless a separate observation infrastructure intervenes. PRFS exploits
that difference: rather than infer the rejected-side outcome from
selection-corrected observations of the accepted side, it measures the
market state directly on the same live venue that produced the
rejection. What is measured is the observed post-rejection market
trajectory, not a Rubin-Neyman potential outcome of a hypothetical
acceptance; the two differ whenever a hypothetical order would have
moved the market, altered fill probability, or triggered execution-model
contingencies. PRFS does not solve the classical selective-labels
identification problem; it side-steps it in the specific DEX setting
where a direct observation of the rejected-side market state is
available, and it names the residual difference between an observational
trajectory and a potential outcome as a matter for the estimand
definition in Section III.G.

\subsection{II.B Backtest selection bias and cadence-matched
evaluation}\label{ii.b-backtest-selection-bias-and-cadence-matched-evaluation}

A parallel concern in quantitative trading is backtest overfitting and
selection bias, where the reported performance of a strategy on a
historical archive overstates its generalisation because the strategy
was selected on that same archive. Bailey and Lopez de Prado (2014)
define the deflated Sharpe ratio as a selection-adjusted performance
measure that accounts for the number of strategy variants tested and the
non-normality of returns. PRFS does not solve backtest selection bias,
but it addresses an adjacent problem: even a rejected candidate's
forward outcome is not available in a typical backtest archive with the
sampling cadence a live decision system uses. Historical price-feed
archives usually store snapshots at fixed calendar intervals
(per-minute, per-hour, per-day) that do not align with the
event-timestamp cadence of a live scanner; on high-turnover DEX
sub-venues the mismatch produces systematic missingness that a backtest
reconstruction cannot recover. A PRFS-style deposit provides a
cadence-matched observational base from which per-filter behaviour can
be characterised without a backtest-side reconstruction, and downstream
backtest evaluations can then reference that base rather than a
mismatched archive.

\subsection{II.C DEX microstructure and empirical DEX
studies}\label{ii.c-dex-microstructure-and-empirical-dex-studies}

Daian et al.~(2020) document maximal extractable value (MEV) and
priority-fee frontrunning in DEX execution, and their taxonomy of
front-running, back-running, and sandwich attacks has been influential
in the DeFi security literature. That literature concerns realised
outcomes of trades that did execute, not observed forward trajectories
of rejected candidates, and the methodological problem addressed by PRFS
is orthogonal to it. Fang et al.~(2022) survey cryptocurrency-trading
empirical methods across both centralised and decentralised venues and
confirm that historical-archive backtesting remains the dominant
approach; their survey does not identify any paper that deposits a
cadence-matched follow-up stream tied to a live rejecting scanner.
Mancino (2025) documents the pump.fun launch environment on Solana,
including the bonding-curve mechanics that produce the token population
studied here; that reference grounds the institutional description of
Section IV and clarifies why the deposited window carries a distinctive
launch-dynamics signature.

\subsection{II.D Positioning}\label{ii.d-positioning}

PRFS is a domain-specific operationalisation of the selective-labels
observation problem for a live DEX filter stack. Its methodological
novelty is not the abstract idea of logging rejected instances (that
idea is present, informally, in operational practice and in the
selective-labels literature). Its novelty is the systematic composition
of (i) a scheduled-cadence sampler tied to the same live oracle path as
the rejecting scanner, (ii) an interval-censored
disappearance-terminated observation contract, (iii) a
reason-attribution rule that identifies the specific filter responsible
for each rejection, and (iv) a public, byte-hashed, code-reproducible
deposit that lets any reader regenerate every reported number from the
frozen CSVs. Sections III through V make each of these four components
explicit and Section VIII shows what the resulting measurement stream
looks like on a specific deposit.

\section{III. Measurement Problem and PRFS
Framework}\label{iii.-measurement-problem-and-prfs-framework}

\subsection{III.A The measurement
problem}\label{iii.a-the-measurement-problem}

Fix a filter-gated algorithmic trading system that evaluates a stream of
candidate tokens and, for each candidate, produces either an accept
decision (the token is admitted for downstream position sizing and
execution) or a reject decision carrying a categorical reason (the token
failed at least one filter). For each accept decision, the system
observes the realised outcome of the position it opens: the entry price,
the subsequent trajectory, the exit price, and the profit-and-loss. For
each reject decision, the system observes nothing beyond the rejection
event itself. Any evaluation of filter behaviour against realised market
outcomes therefore requires an observation of the rejected side of the
decision. The measurement problem is to obtain that observation on the
same live venue that produced the rejection, at a cadence comparable to
the live decision cadence, without perturbing the rejecting scanner. The
observation must survive three practical obstacles: the token can
disappear from the live oracle used by the scanner (short-lifecycle
launch tokens are a common case in the pump.fun setting), the sampler
can lose the oracle connection transiently (persistent failure must be
distinguished from transient failure), and the reason label associated
with the rejection can drift between the point at which it is emitted by
the scanner and the point at which the sampler records it (attribution
must survive the round trip). PRFS is the design that responds to these
three obstacles.

\subsection{III.B PRFS as a measurement
primitive}\label{iii.b-prfs-as-a-measurement-primitive}

PRFS is defined as a triple of interacting subsystems that sit beside
the rejecting scanner. The rejection stream is a durable append-only
event log written by the scanner at the moment of each rejection,
carrying fields (mint, timestamp, reason, p0), where mint is the token
identifier, timestamp is the epoch-millisecond of the rejection, reason
is the categorical label of the filter that fired, and p0 is the venue
mid at the tick's timestamp (FORMAL\_SPEC section 3.1). The scheduler
subscribes to the rejection stream and, for each rejection event,
enqueues follow-up sample requests at pre-specified time offsets. The
oracle adapter executes each scheduled request against the same live
price-and-liquidity oracle used by the rejecting scanner, and writes the
result to a durable append-only outcome log carrying the four registry
fields plus the sample timestamp, sample price, and sample liquidity.
The three subsystems share no in-memory state and communicate only
through the durable logs; the scheduler can restart independently of the
scanner without loss of pending sample requests, and the oracle adapter
can be swapped for an alternative implementation without changing the
scheduler or the scanner. The reference implementation prfs v2.0.0
packages the scheduler and the oracle adapter as a single Python module
(prfs) and provides adapter interfaces for consumers who intend to
substitute their own oracle back-end.

\subsection{III.C Observation window}\label{iii.c-observation-window}

The observation window is the closed time interval from the rejection
timestamp to the analytic follow-up horizon DELTA\_MAX\_MS = 743,040,000
milliseconds, which is 8.60 days exactly (FORMAL\_SPEC section 2).
Samples scheduled past the collection cutoff T\_cut are right-censored;
those with no scheduled tick observed but a neighbouring adaptive tick
within Delta\_probe are interval-censored. The censoring indicator is
emitted by prfs.censoring.label\_censoring on every emitted row so that
downstream consumers can distinguish the three cases: a matched sample,
an interval-censored gap, and a right-censored horizon overrun. The
horizon choice trades statistical resolution against operational cost:
longer horizons capture more of the token's trajectory but require the
sampler to hold pending state for longer and to compete with newer
rejections for scheduler bandwidth. The 8.60-day value used here
reflects the operator's original production cadence and is not a claim
about the correct choice for other deposits.

\subsection{III.D Sampling schedule}\label{iii.d-sampling-schedule}

The reference-implementation default schedule is Schedule\_fix = \{5,
15, 60, 240, 1440\} minutes from the rejection timestamp (FORMAL\_SPEC
section 3.2). The five ticks are chosen to characterise short-horizon
dynamics (5 and 15 minutes capture the immediate post-rejection
response), intermediate-horizon dynamics (60 and 240 minutes capture the
intra-day trajectory), and end-of-day dynamics (1440 minutes captures
the 24-hour follow-up). An adaptive schedule (FORMAL\_SPEC section 3.3)
may add extra ticks driven by log-price or log-liquidity moves exceeding
per-side thresholds, so that a sudden burst of activity on a rejected
token produces additional samples even between the fixed ticks. The
deposit reflects the fixed schedule realised under production tracker
constraints; the adaptive extras are supported by the reference
implementation but were not enabled during the deposited window.

\subsection{III.E Observation predicate and
retry}\label{iii.e-observation-predicate-and-retry}

A scheduled sample is successful if the oracle returns a validated
response with priceUsd greater than zero, liquidity at least zero,
volume24h at least zero, and a non-null pairAddress (FORMAL\_SPEC
section 3.4). A failed query is retried up to R\_max times with jittered
exponential backoff (FORMAL\_SPEC section 3.5); persistent failure after
R\_max retries is recorded as an oracle-absence placeholder rather than
silently dropped, so the sample-side log preserves the fact that a
scheduled sample was attempted and did not succeed. Tokens that
disappear from the oracle before their horizon completes terminate the
sampling loop for that event; the disappearance is recorded as a
terminal marker so that downstream analytics can distinguish an
incomplete sample from an actively terminated one.

\subsection{III.F Reason-attribution
rule}\label{iii.f-reason-attribution-rule}

The reason of a rejection has two sources on disk (FORMAL\_SPEC section
5): reason\_reg (from rejections.csv, written at emission time) and
reason\_out (the copy carried on every sample row in
rejection\_outcomes.csv). The rule (FORMAL\_SPEC section 5.2) is that
reason\_reg is authoritative for per-filter partitions. Every (mint,
timestamp) where the set of distinct reasons on the registry side
differs from the set of distinct reasons on the outcome side is entered
in REASON\_CONFLICT\_LEDGER.csv with disposition
KEEP\_REGISTRY\_AS\_PRIMARY and a two-level confidence label. Within-key
duplicate same-reason sample rows on the outcome side (scheduled-sample
repetitions of one logical attribution) are collapsed under set
semantics; the intent is enforced by
tests/test\_reason\_ledger.py::test\_reason\_set\_semantics\_within\_key\_duplicates\_collapse.
No source reason is silently rewritten. This is a deliberate design
choice: rather than overwrite one side with the other, PRFS preserves
both and marks the disagreement, so that downstream analysts can decide
for themselves whether to trust the registry side, the outcome side, or
neither.

Timestamp identity throughout the coverage and reason-attribution
pipeline is decided on the canonical epoch-millisecond form (via
prfs.clock.to\_epoch\_ms), not on the raw ISO literal; the raw literal
is preserved on every emitted row for byte-provenance audit. On this
deposit both forms agree because every literal already has the shape
\%Y-\%m-\%dT\%H:\%M:\%S.\%fZ; the canonical comparison guards the
pipeline against future mixed-precision inputs.

\subsection{III.G Estimand and scope}\label{iii.g-estimand-and-scope}

Two candidate estimands are defined (FORMAL\_SPEC section 6). The
default is the observed post-rejection forward return r(e, tau) =
(obs\_price(e, tau) / p0(e)) - 1 restricted to e.reason = f\_i for the
per-filter distribution. This is an observational quantity computed from
the observed forward market trajectory of the specific rejected mint; it
is not a Rubin-Neyman potential outcome of a hypothetical acceptance,
because a hypothetical order would have introduced price impact, fill
uncertainty, and execution-model contingencies that the observed
trajectory does not capture. A hypothetical executed-PnL estimand is
defined but requires an explicit ExecutionModel (entry slippage, exit
slippage, fees, exit rule, portfolio constraint) and the reference
implementation raises prfs.estimand.MissingExecutionModel otherwise.
This paper reports only r(e, tau) and does not compute or claim any
executed profit-and-loss figure. The design is observational and the
statistics are descriptive. The distinction matters because the
interpretation of a coverage figure depends on it: 48.55 percent primary
event coverage means 48.55 percent of the deposit's 2,997 rejections
have a matched observed forward trajectory in the deposit; it does not
mean 48.55 percent of a hypothetical acceptance distribution has been
observed.

\section{IV. Data and Institutional
Setting}\label{iv.-data-and-institutional-setting}

\subsection{IV.A Deposited dataset and
provenance}\label{iv.a-deposited-dataset-and-provenance}

The deposited dataset (Zenodo concept DOI 10.5281/zenodo.20043515)
contains two CSV files whose SHA-256 values match the deposited
checksums.txt (verified in the reproducibility package under data/):

\begin{table}[!htbp]
\centering
\small
\begin{tabular}{l S[table-format=6.0] l}
\toprule
{\bfseries File} & {\bfseries Rows} & {\bfseries SHA-256 (first 16 hex)} \\
\midrule
\texttt{rejections.csv}          & 2997  & \texttt{2ccb11d23d3a673a} \\
\texttt{rejection\_outcomes.csv} & 67000 & \texttt{7bcb0eb2a136b611} \\
\bottomrule
\end{tabular}
\end{table}

The deposit is version-locked: any consumer who obtains the two files
with the checksums.txt file present can verify the byte identity of
every row before running any reproducibility script. The concept DOI
resolves to the latest version of the deposit and is the citation form
recommended for downstream references.

\subsection{IV.B Observation environment (Solana pump.fun
sub-venue)}\label{iv.b-observation-environment-solana-pump.fun-sub-venue}

The deposited data were collected on a single Solana
automated-market-maker sub-venue that surfaces bonding-curve launches
originating on the pump.fun platform (see Mancino 2025 for a general
description of that launch environment). The rejecting scanner and the
PRFS oracle adapter both queried the same public DEX-aggregator feed for
tokens that the scanner had evaluated as candidates in the pump.fun
launch pipeline. This scope has three implications for the
interpretation of the deposited numbers. First, the token population is
dominated by short-lifecycle launch tokens whose oracle presence can end
abruptly, which shapes the interval-censoring pattern documented in
Section VIII. Second, the coverage figures reported here are per-event
coverage on a single sub-venue during a specific launch-dynamics window;
they should not be read as generalising to other AMM DEX venues, to
Ethereum layer-2 surfaces, or to non-launch-focused trading systems
without an explicit replication on those settings. Third, the
microstructure literature on maximal extractable value and priority-fee
frontrunning (Daian et al.~2020) is orthogonal to the measurement
question addressed here but is relevant background for any reader
interpreting the executed side of the trading system's decisions; PRFS
deliberately restricts itself to the observation side and does not
intervene in the execution path.

\subsection{IV.C Observation window and calendar
span}\label{iv.c-observation-window-and-calendar-span}

The registry timestamps span 2026-04-10T21:10:13.835Z through
2026-04-19T12:12:13.333Z UTC. The observed calendar span (max minus min
of the registry timestamps) is 8.626 days, rounded to two decimal places
8.63 days. This is distinct from the analytic follow-up window
DELTA\_MAX\_MS = 743,040,000 milliseconds = 8.60 days exactly defined in
FORMAL\_SPEC section 2; the window governs the scheduled-sample cadence,
the calendar span is a descriptive property of the deposit.

\subsection{IV.D Deposit composition}\label{iv.d-deposit-composition}

The deposit contains 457 unique mints and seven anonymised filter labels
filter\_1 through filter\_7 (numeric ordering is label-anonymised; it
does not correspond to volume or rule identity). The scanner emitted
rejections under one of two source labels, source\_a and source\_b,
which are documented in Section V.B; the two labels denote different
upstream input topologies of the same scanner stack, not two
independently maintained scanner codebases. The registry emits
log2-quantised liquidity and volume bins alongside the raw values; the
anonymisation and quantisation are documented in the deposit README.
This composition matters for interpretation: because filter\_2 alone
contains 48 percent of the deposit and because the two source topologies
partition roughly evenly, per-filter and per-source analyses cannot
treat the categories as equal-population strata, and Section VIII
reports the per-category denominators alongside every point estimate so
that comparative claims can be evaluated against the correct base.

\section{V. System Architecture and
Implementation}\label{v.-system-architecture-and-implementation}

\subsection{V.A Reference implementation versus production
system}\label{v.a-reference-implementation-versus-production-system}

The reference implementation deposited alongside this manuscript (prfs
v2.0.0, under IMPLEMENTATION/prfs/) is a specification-level restatement
of PRFS: it is the code path against which every headline number in this
manuscript is regenerated under continuous integration, and it
implements the formal specification (FORMAL\_SPEC.md) end-to-end. It is
not the production system that emitted the deposited CSVs. The
production system is an operational scanner-plus-tracker stack developed
and operated separately, whose implementation details are not part of
this paper's contribution. What the reference implementation does
guarantee is that any consumer of the deposit can regenerate every
reported number from the byte-identical CSVs using only the deposited
code, without access to the production system. A distinguishing sentence
to this effect is included in the deposit README. The consequence for
interpretation is that restart claims, WAL correctness,
reason-attribution behaviour, and cross-check agreement are validated on
the reference implementation; they are consistent with the deposit but
do not certify any specific behaviour of the production system that
produced the deposit.

\subsection{V.B Scanner and source
labels}\label{v.b-scanner-and-source-labels}

The scanner ingests two input topologies. Source label source\_a
corresponds to a lower-latency direct-launch feed; source label
source\_b corresponds to a higher-latency aggregated feed. Both feed
into the same filter chain and both emit into the same registry. The two
labels are preserved on every registry row and their differential
coverage behaviour is quantified in Section VIII. The phrase ``single
scanner stack'' in the deposit README describes the shared
filter-and-emission code path; the two source labels are a topology
diagnostic within that shared stack, not two separately maintained
scanners. A consequence of this design is that any per-filter analysis
on the deposit should also condition on source, because the joint
distribution of source and filter across events is not uniform; Section
VIII.D quantifies the resulting source-associated missingness.

\subsection{V.C Persistence and restart
discipline}\label{v.c-persistence-and-restart-discipline}

The registry and the outcome log are both append-only write-ahead logs
(prfs.persistence.WalEventStore and WalSampleLog) with a mark\_terminal
call that pins a specific event as fully processed. On
operator-initiated restart, prfs.persistence.restart\_recover replays
events and samples from the WAL tails and re-enqueues any pending sample
requests that had not been marked terminal at shutdown. Restart claims
are validated by tests/test\_restart.py against a deterministic
simulator that injects restart events at controlled points in the sample
schedule and verifies that the resulting outcome log is bit-identical to
the log produced under uninterrupted operation. This is not a proof of
production-system restart correctness (the production system uses its
own persistence layer) but it is a proof that the reference
implementation, which is the code path any consumer of the deposit will
run, is deterministic under restart.

\subsection{V.D Cross-check
aggregator}\label{v.d-cross-check-aggregator}

To catch code-path bugs in the primary reference implementation, a
secondary cross-check aggregator (oracle/oracle\_aggregate.py) was
built. It uses csv.reader with explicit column-index lookup and a
different accumulation strategy than the primary prfs.coverage and
prfs.reason\_ledger code path, and does not import either module. It
shares the same logical set-intersection framework and the same Python
runtime, so agreement between the two implementations is
cross-verification of the reference implementation, not evidence of
language-independent verification. On the deposited files it reproduces
every headline number identically: 1,455 / 2,997 primary, 1,641 / 2,997
secondary, 253 conflict emissions, 263 orphan keys, and the same seven
per-filter rows. The pytest suite's test\_oracle\_agreement.py gates on
this agreement, so any code change to either implementation that breaks
the agreement fails the test suite. A third independent stdlib-only
recompute script (scratchpad/p2\_repro/independent\_repro.py) that
imports neither prfs nor oracle also matches every headline number and
is included in the reproducibility package for readers who wish to
verify without executing either the reference implementation or the
cross-check aggregator.

\section{VI. Matching and Outcome
Construction}\label{vi.-matching-and-outcome-construction}

\subsection{VI.A Two candidate keys}\label{vi.a-two-candidate-keys}

Two candidate event keys are defined and computed by the reference
implementation. The primary key is reason-aware and three-field: (mint,
timestamp, reason) on the registry side, matched against (mint,
rejectTs, rejectReason) on the outcome side. The secondary key is
reason-agnostic and two-field: (mint, timestamp) matched against (mint,
rejectTs). Both matches are computed as exact string equality after
canonical epoch-millisecond conversion; on this deposit every raw
timestamp already has millisecond precision in the shape
\%Y-\%m-\%dT\%H:\%M:\%S.\%fZ, so the canonical form and the raw literal
agree bit for bit.

\subsection{VI.B Why the three-field key is
primary}\label{vi.b-why-the-three-field-key-is-primary}

The three-field key treats each rejection event as a (candidate,
moment-in-time, rule-that-fired) triple. That is the correct unit of
analysis for filter attribution, because any downstream per-filter
statistic asks: for the events attributed to filter X, what is the
observed post-rejection outcome distribution? Attribution requires the
reason label to survive from registry to outcome. If the outcome side
records a different reason at the same (mint, timestamp), the event
either was subject to downstream re-labelling or was a co-firing
artefact where multiple rules fired on the same evaluation and only one
made it to the outcome row. Either way, an event lacking a shared reason
label between the two sides cannot be counted toward a specific-filter
coverage figure without an additional assumption. The reason-aware
three-field key is therefore the identified estimand for per-filter
analyses and is the primary headline throughout this paper.

\subsection{VI.C Why the two-field key is diagnostic
only}\label{vi.c-why-the-two-field-key-is-diagnostic-only}

The two-field key answers a strictly weaker question: did the sampler
ever fire on this candidate at this moment, regardless of which filter
had rejected it? That number is scientifically meaningful as an
integrity check on the sampler itself: it isolates the question of
whether the scheduler and oracle adapter reached the token from the
question of whether the reason attribution survived. Because it is
silent about attribution, the two-field number is reported here only as
a diagnostic and is not used for any per-filter analysis. The diagnostic
surplus (secondary minus primary = 1,641 - 1,455 = 186 events) measures
the volume of (mint, timestamp) matches where the reason changed between
the two sides; the reason-conflict ledger of Section VIII accounts for
these.

\subsection{VI.D Outcome construction}\label{vi.d-outcome-construction}

For each matched event, the per-tau observed forward return is r(e, tau)
= (obs\_price(e, tau) / p0(e)) - 1 with obs\_price(e, tau) taken from
the outcome row whose sample timestamp is closest to (rejectTs + tau)
within a tolerance defined by prfs.censoring.label\_censoring. When no
outcome row falls within the tolerance for a given tau, the
corresponding r(e, tau) is marked missing and the censoring indicator
records whether the miss is a right-censor (past horizon), an
interval-censor (adjacent scheduled tick within Delta\_probe), or a
persistent oracle absence. Downstream statistics distinguish the three
miss categories rather than pooling them; the deposit ships with a
censoring-summary CSV whose per-filter distributions are the input to
any survival-style analysis a consumer may wish to run.

\section{VII. Statistical Methods}\label{vii.-statistical-methods}

\subsection{VII.A Confidence intervals for
coverage}\label{vii.a-confidence-intervals-for-coverage}

For each coverage proportion the paper reports the two-sided 95 percent
Wilson-score binomial confidence interval (Wilson 1927). Wilson
intervals are preferred to the normal approximation because they retain
nominal coverage at small numerators or denominators; small-n categories
(Wilson interval width above 30 percentage points) are flagged with a
sparse-category warning, and per-category point estimates within those
categories are treated as unstable. No p-values or hypothesis tests are
reported on the headline coverage numbers; the paper is a measurement
contribution, not an inferential one. Two exceptions to the no-p-values
rule are reported explicitly as evidence against the
missing-completely-at-random hypothesis: a Pearson chi-square test on
the 2-by-7 filter-versus-coverage contingency table (Section VIII.C) and
a two-proportion z-test on the source\_a versus source\_b coverage
difference (Section VIII.D). Both are reported as tests of the MCAR
null, not as tests of any filter-level or source-level effect.

\subsection{VII.B Event-independence caveat and
clustering}\label{vii.b-event-independence-caveat-and-clustering}

Wilson intervals treat each event as an independent Bernoulli trial. On
this deposit the assumption is violated in a specific, quantified way:
the 2,997 events are drawn from only 457 unique mints, with mean 6.56
events per mint, median 3, ninetieth percentile 14, and maximum 89; 96
percent of events belong to multi-event mints. Within a mint, coverage
outcomes are almost certainly correlated (the scheduler's success on one
event for a mint is informative about its success on later events for
the same mint, because the same oracle-presence and sample-scheduling
state persists across events on the same token), so the effective sample
size is bounded above by a quantity closer to the number of clusters
(457) than to the number of events (2,997). Under a naive-independence
Wilson interval the aggregate primary coverage 48.55 percent has a 95
percent CI of approximately {[}46.8 percent, 50.3 percent{]}; a
mint-level bootstrap or a mint-cluster-robust interval would be wider.
Section IX reports a mint-level aggregation as a sensitivity check. Any
downstream inference that treats event-level coverage as ignorable,
including but not limited to Kamat 2026a and Kamat 2026b, must respect
this clustering and use cluster-robust standard errors or mint-level
bootstrapping.

The reader should treat the naive-independence Wilson widths reported
throughout Section VIII as lower bounds on the true uncertainty. The
scale of the understatement depends on the intra-mint correlation
structure, which the deposit does not fix a priori; the mint-level
aggregation in Section IX.B provides one way to bound the understatement
in the aggregate, but per-filter and per-day understatement can be
larger or smaller than the aggregate figure depending on how the mints
partition into filters and days.

\subsection{VII.C Missingness}\label{vii.c-missingness}

Missingness is treated as observed and enumerated, not imputed. Where
the missingness is systematically associated with a documented
characteristic (filter identity, source label, day-of-window), the
association is reported (Section VIII) so that any downstream
conditional-independence assumption can be evaluated against the
observed pattern. The paper does not claim missing-completely-at-random
(MCAR); the Section VIII association tests decisively reject MCAR on
this deposit. Nor does the paper claim missing-at-random (MAR) after
conditioning on the observed characteristics: coverage may still depend
on unobserved token features (liquidity depth at rejection, oracle
presence at rejection, downstream oracle disappearance) that the deposit
does not carry. The safest interpretation of the coverage figures is
that they describe the deposit as constructed, and that any
generalisation beyond the deposit requires an explicit assumption about
the missingness mechanism that the deposit does not itself supply.

\section{VIII. Results}\label{viii.-results}

\subsection{VIII.A Primary event coverage (reason-aware three-field
key)}\label{viii.a-primary-event-coverage-reason-aware-three-field-key}

Under the primary three-field key (mint, timestamp, reason), event
coverage on the deposit is:

\begin{itemize}
\tightlist
\item
  matched\_events = 1,455
\item
  denominator = 2,997
\item
  coverage = 48.55 percent
\item
  Wilson 95 percent CI (naive event-independence) = {[}46.76 percent,
  50.34 percent{]}
\end{itemize}

This is the headline number for any per-filter analysis (FORMAL\_SPEC
section 4.1). The interval is reported under the caveat in Section
VII.B: because 2,997 events cluster within 457 mints, the effective
sample size is below 2,997 and the naive interval understates
uncertainty.

\subsection{VIII.B Secondary two-field
diagnostic}\label{viii.b-secondary-two-field-diagnostic}

Under the reason-agnostic two-field key (mint, timestamp), the
diagnostic count is 1,641 matches (54.75 percent), with a naive Wilson
95 percent CI of {[}52.97 percent, 56.53 percent{]}. The 186-event gap
between the primary and diagnostic numbers is not clock skew or
timestamp rounding; both files share millisecond-precision string
timestamps and the pipeline canonicalises to epoch-milliseconds before
comparison, so the gap is a reason-field integrity mismatch across the
two files (FORMAL\_SPEC section 4.2). The two-field number is reported
only as a diagnostic; downstream per-filter analyses always use the
three-field key.

\subsection{VIII.C Per-filter coverage with Wilson binomial confidence
intervals}\label{viii.c-per-filter-coverage-with-wilson-binomial-confidence-intervals}

Coverage is highly non-uniform across filters (see Table 1). Small-n
categories (Wilson CI width above 30 percentage points) are flagged with
a sparse-category warning to signal that per-category point estimates
are unstable and comparative claims between categories should be
interpreted accordingly.

\begin{table}[!htbp]
\centering
\small
\begin{tabular}{l S[table-format=4.0] S[table-format=3.0] r l r c}
\toprule
{\bfseries Reason} & {\bfseries Registry events} & {\bfseries Matched} & {\bfseries Coverage} & {\bfseries 95 percent CI [lo, hi]} & {\bfseries Width} & {\bfseries Sparse?} \\
\midrule
\texttt{filter\_1} &   804 & 579 & 72.01\% & [68.81\%, 75.01\%] &  6.19 pp & no  \\
\texttt{filter\_2} & 1444  & 474 & 32.83\% & [30.45\%, 35.29\%] &  4.84 pp & no  \\
\texttt{filter\_3} &   313 & 219 & 69.97\% & [64.67\%, 74.78\%] & 10.11 pp & no  \\
\texttt{filter\_4} &   317 & 117 & 36.91\% & [31.78\%, 42.35\%] & 10.57 pp & no  \\
\texttt{filter\_5} &    62 &  39 & 62.90\% & [50.46\%, 73.84\%] & 23.38 pp & no  \\
\midrule
\texttt{filter\_6} &    15 &   9 & 60.00\% & [35.75\%, 80.18\%] & 44.43 pp & YES \\
\texttt{filter\_7} &    42 &  18 & 42.86\% & [29.12\%, 57.79\%] & 28.68 pp & no  \\
\bottomrule
\end{tabular}
\caption*{Table 1. Per-filter event coverage under the primary three-field key.}
\end{table}

filter\_2 (48 percent of the deposit) has the lowest coverage at 32.83
percent; filter\_1 has the highest at 72.01 percent. The aggregate 48.55
percent masks this heterogeneity. filter\_6 (n = 15) carries a
sparse-category warning; a coverage difference of tens of percentage
points between it and any other filter is inside its own confidence
interval and should not be interpreted as evidence of a filter-level
effect. The filter-association chi-square on the 2 by 7 coverage table
is chi-square(6) = 401.29 with p approximately zero, so the observed
missingness is not consistent with MCAR under filter.

\begin{figure}
\centering
\includegraphics[width=0.85\linewidth]{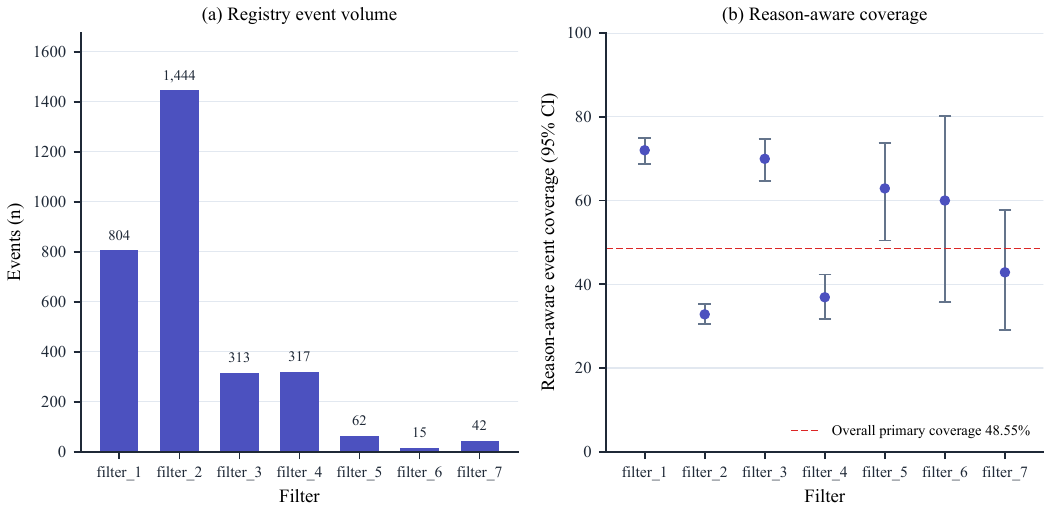}
\caption{Figure 3. Per-filter registry event volume and matched-event
overlay. Grey bars: registry events per filter label (sums to 2,997).
Blue bars: registry events per filter label that match at least one
outcome row on the three-field key (sums to 1,455). Percentage labels
above each blue bar are the per-filter coverage rates from Table 1.
Regenerated deterministically from the frozen deposit.}
\end{figure}

\subsection{VIII.D Source-associated
missingness}\label{viii.d-source-associated-missingness}

Coverage by scanner source label:

\begin{itemize}
\tightlist
\item
  source\_a: 906 matched of 1,450 registry events, 62.48 percent, Wilson
  95 percent CI {[}59.96 percent, 64.94 percent{]}
\item
  source\_b: 549 matched of 1,547 registry events, 35.49 percent, Wilson
  95 percent CI {[}33.14 percent, 37.91 percent{]}
\end{itemize}

The two-proportion z-statistic on source\_a versus source\_b is z =
14.78 with p approximately zero. Coverage is source-associated at
overwhelming significance. Since Section V.B established that source\_a
and source\_b are two upstream input topologies of the same
filter-and-emission code path, the coverage differential is attributable
to topology (feed latency, availability, scheduling order) rather than
to two separately maintained scanner codebases. Any downstream analysis
conditioning on filter should also condition on source, because the
source and the filter distribution across events are not independent.

\subsection{VIII.E First-day warm-up
transient}\label{viii.e-first-day-warm-up-transient}

Per-day event coverage under the primary three-field key is shown in
Table 2.

\begin{table}[!htbp]
\centering
\small
\begin{tabular}{l S[table-format=3.0] S[table-format=3.0] r}
\toprule
{\bfseries Day (UTC)} & {\bfseries Registry events} & {\bfseries Matched (3-field)} & {\bfseries Coverage} \\
\midrule
2026-04-10 (partial from 21:10Z)    & 148 &   2 &  1.35\% \\
2026-04-11                          & 626 &  57 &  9.11\% \\
\midrule
2026-04-12                          & 256 & 178 & 69.53\% \\
2026-04-13                          & 338 & 205 & 60.65\% \\
2026-04-14                          & 439 & 248 & 56.49\% \\
2026-04-15                          & 339 & 236 & 69.62\% \\
2026-04-16                          & 184 & 126 & 68.48\% \\
2026-04-17                          & 273 & 170 & 62.27\% \\
2026-04-18                          & 231 & 143 & 61.90\% \\
2026-04-19 (partial through 12:12Z) & 163 &  90 & 55.21\% \\
\bottomrule
\end{tabular}
\caption*{Table 2. Per-day event coverage.}
\end{table}

The earliest sampleTs in rejection\_outcomes.csv is
2026-04-11T20:26:53.379Z, which is 23 hours 17 minutes after the
earliest rejection in the deposit at 2026-04-10T21:10:13.835Z; its own
rejectTs is 2026-04-11T20:21:42.531Z, so it is a normal 5-minute
scheduled tick for a subsequent rejection of the same mint. No sample
in the deposit is directly associated with the first rejection: if the
sampler had been running with the fixed cadence \{5, 15, 60, 240, 1440\}
minutes from the first rejection onward, a sample at rejection + 5
minutes would exist at 2026-04-10T21:15:13Z; no such row appears in the
deposit. Two interpretations are consistent with the data: either the
sampler warmed up during the first partial calendar day and did not
begin emitting sample rows until 2026-04-11T20:26Z, or the deposit was
truncated on the sample side and pre-2026-04-11T20:26Z sample rows were
dropped before deposit. The deposit itself does not distinguish these;
both are disclosed here. Section IX reports a sensitivity check that
excludes the first 26.8 hours of registry events, and Section XI names
this transient in the external-validity discussion. The 48.55 percent
aggregate primary coverage is therefore a mixture of a warm-up regime
(approximately 7.6 percent coverage on the first 26.8 hours, comprising
the partial 2026-04-10 UTC day from 21:10Z onward plus the full
2026-04-11 UTC day, 774 events, 59 matched) and a steady-state regime (62.8
percent coverage from 2026-04-12 onward, 2,223 events, 1,396 matched).

\subsection{VIII.F Mint-level
coverage}\label{viii.f-mint-level-coverage}

All 457 unique mints in rejections.csv appear somewhere in
rejection\_outcomes.csv (100 percent, Wilson 95 percent CI {[}99.17
percent, 100.00 percent{]}). Mint-level coverage is a coarser unit than
event coverage and cannot be substituted for the per-event or per-filter
numbers. The 100 percent mint coverage is an arithmetic consequence of
clustering: mints are heavily repeated (mean 6.56 events per mint, 72
percent of mints in more than one rejection event), so a mint appears in
the outcome file whenever the sampler ever fires on any one of its
events. Under those clustering statistics, 100 percent mint coverage is
compatible with any event coverage above 15.2 percent (that is, 457 /
2,997); the observed 48.55 percent event coverage has 100 percent mint
coverage as an emergent property, not an independent achievement. The
reader who intends to cite the 100 percent number as evidence of sampler
completeness should therefore note the distinction: mint-level coverage
says the sampler reached every rejected token at least once, and
event-level coverage says how often it reached each rejection event on
that token.

\begin{figure}
\centering
\includegraphics[width=0.85\linewidth]{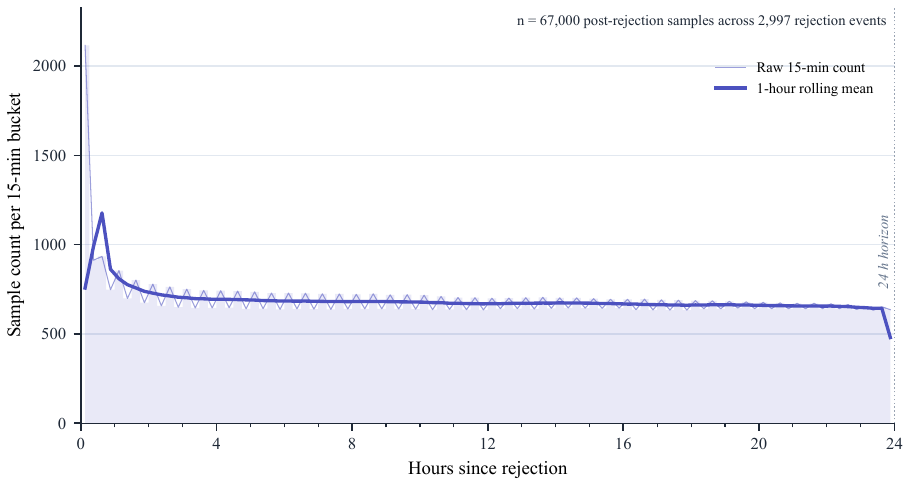}
\caption{Figure 2. Per-event forward-sample-density histogram. For each
distinct outcome three-field key (mint, rejectTs, rejectReason) the
number of scheduled sample rows is counted; the grey histogram covers
all distinct outcome three-field keys, and the blue overlay covers the
subset that match a registry event on the primary three-field key. The
grey excess above the blue bars corresponds to the orphan-key population
documented in Section VIII.H. Regenerated deterministically from the
frozen deposit.}
\end{figure}

\subsection{VIII.G Sample-level alignment with dual
denominators}\label{viii.g-sample-level-alignment-with-dual-denominators}

58,407 outcome rows attach to a matched three-field event key. Reported
against both documented denominators:

\begin{itemize}
\tightlist
\item
  Against the full outcome row count of 67,000: 58,407 / 67,000 = 87.17
  percent.
\item
  Against the eligible denominator (67,000 minus the 8,593 outcome rows
  that attach to one of the 263 orphan three-field outcome keys, that
  is, keys that share (mint, rejectTs) with a registry row under a
  different reason and therefore cannot align by construction): 58,407 /
  58,407 = 100.00 percent.
\end{itemize}

Both denominators are surfaced in every downstream artefact; consumers
of the number must state which denominator is being cited.

\subsection{VIII.H Reason-conflict
ledger}\label{viii.h-reason-conflict-ledger}

REASON\_CONFLICT\_LEDGER.csv contains 253 rows, one per (mint,
timestamp) where the set of distinct reasons recorded on the registry
side differs from the set of distinct reasons recorded on the outcome
side. Sum of orphan\_key\_count across the ledger is 263, matching
exactly the number of orphan three-field outcome keys computed in the
primary coverage pass. Every row's default disposition is
KEEP\_REGISTRY\_AS\_PRIMARY with a two-level confidence label (HIGH when
both sides carry reasons the other side lacks, MEDIUM when one side is a
proper subset of the other). Alternative dispositions
(KEEP\_OUTCOME\_AS\_PRIMARY, MULTI\_FILTER\_CO\_FIRING,
ANONYMISATION\_MAP\_DRIFT) are reserved but not applied; no source
evidence in the deposit supports them. Read together with Section
VIII.G, 253 conflict events represent 15.4 percent of the 1,641 events
matched under the two-field diagnostic key: the reason-attribution gap
is non-negligible and is disclosed rather than hidden.

\section{IX. Robustness and
Reproducibility}\label{ix.-robustness-and-reproducibility}

\subsection{IX.A First-day
sensitivity}\label{ix.a-first-day-sensitivity}

Excluding the first 26.8 hours of registry events (2026-04-10T21:10Z
through the end of the 2026-04-11 UTC day, 774 events) from the primary
coverage computation raises the aggregate primary coverage from 48.55
percent (1,455 / 2,997) to 62.8 percent (1,396 / 2,223) on the
steady-state subset. This does not change the paper's headline (the headline is the
deposit-wide number for reproducibility with the full deposit) but
establishes the sensitivity: readers who intend to characterise
steady-state sampler behaviour rather than deposit-wide behaviour should
apply the exclusion and cite the steady-state figure. The full-deposit
and steady-state numbers are both regenerated by the reference
implementation from the frozen CSVs.

\subsection{IX.B Mint-level aggregation as clustering
sensitivity}\label{ix.b-mint-level-aggregation-as-clustering-sensitivity}

As a clustering sensitivity, compute per-mint coverage as p\_mint =
matched\_events\_for\_mint / total\_events\_for\_mint on each of the 457
mints, then take the equal-weighted mean over mints. The mint-level mean
of per-mint coverage is 77.71 percent, with a normal-approximation 95
percent CI on that mean (n = 457) of {[}75.12 percent, 80.30 percent{]}.
This is higher than the event-level 48.55 percent because mints with
many events tend to be persistent tokens the sampler covers well, so
weighting by event pulls the average toward the heavy-tail
highly-rejected mints where the sampler mostly failed. Neither statistic
is the correct one on its own; they answer different questions. The
event-level 48.55 percent answers: for a random rejection event, what is
the probability the deposit shows an outcome? The mint-level 77.71
percent answers: for a random rejected mint, what is the average
within-mint coverage? Per-filter analyses that condition on the
rule-that-fired belong to the event-level question and must respect the
mint clustering with cluster-robust standard errors as noted in Section
VII.B.

\subsection{IX.C End-to-end
regeneration}\label{ix.c-end-to-end-regeneration}

Every reported number in this manuscript is regenerated from the
byte-identical deposited CSVs by python -m prfs.cli -{}-registry-csv
data/rejections.csv -{}-outcomes-csv data/rejection\_outcomes.csv
-{}-report-out reports/verify\_report.json, and is cross-verified by the
secondary aggregator oracle/oracle\_aggregate.py, which does not import
prfs. The pytest suite gates on both. An independent stdlib recompute
(scratchpad/p2\_repro/independent\_repro.py, no prfs imports, no oracle
imports) also matches every headline number and is included in the
reproducibility package for readers who wish to verify without executing
the reference implementation. The three code paths agree on all headline
numbers documented in Section VIII, so any subsequent code change that
breaks one path without breaking the others produces a test-suite
failure that names the disagreement.

\subsection{IX.D What the reproducibility package
contains}\label{ix.d-what-the-reproducibility-package-contains}

The reproducibility package that accompanies the deposit contains: the
two data files rejections.csv and rejection\_outcomes.csv with a
checksums.txt file that lists their SHA-256 values; the reference
implementation prfs v2.0.0 including the coverage, reason-ledger,
censoring, and clock modules; the cross-check aggregator
oracle/oracle\_aggregate.py; the independent stdlib recompute script
scratchpad/p2\_repro/independent\_repro.py; a pytest test suite that
gates on the agreement of the three code paths; a
NUMBER\_PROVENANCE\_FINAL.csv file that maps every headline number in
this manuscript to its formula, its source CSV, and its code location;
and the deposit README that documents the schema of the two CSVs, the
anonymisation of filter and source labels, the window boundaries, and
the concept DOI for citation. A reader who obtains the package can
therefore verify the deposit at three levels: byte-identity (against
checksums.txt), computed-value identity (against
NUMBER\_PROVENANCE\_FINAL.csv), and code-path agreement (via the pytest
suite). No component of the package requires network access at run time.

\section{X. Discussion}\label{x.-discussion}

The measurement problem PRFS addresses is not exotic. Every filter-gated
algorithmic trading system that operates on a live oracle produces a
rejection stream and could, in principle, subscribe a follow-up sampler
to that stream. The question is not whether it is possible to log
rejected tokens (many production systems do, in some form), but whether
the logging is systematic enough to support downstream measurement
claims. Ordinary operational logs typically fail on one or more of five
specifications: they do not schedule samples at a defined cadence, they
do not tie the sample query to the same oracle path used by the scanner,
they do not preserve the reason label through the round trip, they do
not distinguish oracle absence from transient failure, and they do not
produce a public deposit whose numbers a reader can regenerate.

What the deposited artefacts show is that satisfying all five
specifications, at once and on the same live venue that produced the
rejections, materially changes what a downstream analyst can conclude
about filter behaviour. The 48.55 percent primary coverage is not a
target; it is the observed coverage on this specific deposit under this
specific scheduler and oracle. Both the per-filter heterogeneity (32.83
percent to 72.01 percent) and the per-source heterogeneity (35.49
percent to 62.48 percent) are the interesting scientific facts, because
they say the aggregate is a mixture and the mixture components are not
exchangeable. The 253-row reason-conflict ledger and the 8,593 unaligned
outcome rows are the reproducibility contract: they name a residual
attribution gap rather than paper over it, and any downstream user of
the deposit must decide how to treat those rows in their own analysis.

The design deliberately restricts the reported estimand to the observed
post-rejection forward return. Any stronger claim about what a
hypothetical acceptance would have earned requires an explicit
ExecutionModel, because a hypothetical order would have introduced price
impact, fill uncertainty, and portfolio-constraint interactions that the
observed trajectory does not capture. That claim is not made anywhere in
this paper.

Positioned against the selective-labels literature (Section II.A), PRFS
does not resolve the classical identification problem: it does not
construct a hypothetical acceptance outcome, and it does not claim that
its coverage figures identify any latent parameter of the
accepted-versus-rejected decision process. What it does is exploit a
structural feature of the DEX setting (the market state of a rejected
token continues to evolve after the rejection) that is absent from the
classical bail, credit, and hiring settings, and turn that structural
feature into a measurement contract. The output is a base for
observational analyses; downstream inferential work that seeks to
identify a causal parameter of the accepted-versus-rejected decision
must still supply the identifying assumption from outside PRFS.

Positioned against ordinary operational logging, PRFS contributes the
systematic combination of the five specifications above and the explicit
reproducibility contract that ties every reported number back to a
deposited CSV. Operational logs written for post-mortem debugging can,
in principle, support downstream measurement, but in practice they
typically do not, because their design goals (rapid recovery from
operational faults, minimal disk cost, human-readable diagnostics)
diverge from the design goals of a measurement instrument (defined
cadence, oracle-path identity, reason preservation, censoring taxonomy,
byte-hashed deposit). PRFS is not an alternative to operational logging;
it is a separate subsystem with a measurement-instrument design.

\section{XI. Limitations}\label{xi.-limitations}

External validity. The deposit reflects a single Solana
automated-market-maker sub-venue (Solana pump.fun bonding-curve
launches) during an 8.63-calendar-day window in April 2026. The token
population, the launch dynamics, the scheduler cadence, and the oracle
path are all specific to that setting. Coverage figures reported here
should not be read as generalising to other AMM DEX venues, to Ethereum
layer-2 surfaces, to non-launch-focused trading systems, or to different
scheduler cadences without an explicit replication.

Filter and source population. The seven anonymised filter labels reflect
the specific rules deployed in the operator's production filter chain
during the window. Any downstream analyst comparing filters must respect
the per-filter denominators (Table 1) rather than treating the seven
categories as equal-population strata; filter\_2 alone contains 48
percent of the deposit.

Missingness. Missingness is not MCAR: coverage is filter-associated
(chi-square(6) = 401.29, p approximately zero), source-associated (z =
14.78, p approximately zero), and time-associated (per-day coverage
ranges 1.35 percent to 69.62 percent). Any inference that treats
event-level coverage as ignorable must condition on filter, source, and
day-of-window at minimum. Missing-not-at-random cannot be ruled out
either: coverage may also depend on unobserved characteristics of the
token such as liquidity depth at rejection, oracle presence at
rejection, or downstream oracle disappearance.

Warm-up transient. The first 26.8 hours of the deposit (774 events)
carry a distinctive low-coverage regime (7.6 percent) that is
inconsistent with the steady-state coverage (62.8 percent) from
2026-04-12 onward. The deposit itself does not distinguish sampler
warm-up from sample-side truncation; both are disclosed and a
sensitivity check is provided in Section IX.A.

Clustering and independence. Events cluster within mints (mean 6.56
events per mint, 96 percent of events in multi-event mints). Naive
event-level Wilson intervals reported here understate uncertainty by a
factor bounded above by the square root of the design effect; any
downstream inference must use cluster-robust or mint-level bootstrap
standard errors.

Reference implementation versus production system. The reference
implementation prfs v2.0.0 is a specification-level restatement that
regenerates every number in this manuscript. The production system that
emitted the deposited CSVs is a separately operated scanner-plus-tracker
stack. Restart claims, WAL correctness, and reason-attribution behaviour
are validated on the reference implementation; they are consistent with
the deposit but do not certify the production system's implementation.

Small-n categories. filter\_6 (n = 15) has a Wilson 95 percent
confidence interval width of 44.43 percentage points and is flagged with
a sparse-category warning; any comparative claim involving filter\_6
should be read against that interval.

Censoring. Scheduled samples past the collection cutoff T\_cut are
right-censored; samples with no scheduled tick observed but a
neighbouring adaptive tick within Delta\_probe are interval-censored.
The censoring indicator is emitted by prfs.censoring.label\_censoring on
every row.

Downstream external interpretations. Classification rules that treat
oracle disappearance as a positive signal (Kamat 2026a, Kamat 2026b)
belong to those downstream papers and are not validated by the present
paper.

No causal claim. The design is observational. This paper reports
observed counts, calculated proportions, and their binomial uncertainty.
It does not use the words causal, counterfactual, treatment, effect,
impact, or caused by in a claim of identification, and does not claim
identification. Where prior literature (selective labels, reject
inference) uses ``counterfactual'' as its own established term of art,
that usage is preserved by attribution in Section II and is not extended
to PRFS's own estimand.

\section{XII. Conclusion}\label{xii.-conclusion}

Under the reason-aware three-field event key (mint, timestamp, reason),
the deposited PRFS dataset achieves 1,455 of 2,997 (48.55 percent)
primary event coverage; under the reason-agnostic two-field diagnostic
key, 1,641 of 2,997 (54.75 percent). This paper supplies the formal
specification, the reference implementation, and the secondary
cross-check implementation that let any reader regenerate every number
from the byte-identical deposited CSVs. Coverage is highly non-uniform
across filter categories, scanner sources, and days of the observation
window; the 253-row conflict ledger and 263 orphan keys quantify the
residual reason-field integrity gap; and the deposit-wide aggregate is
shown to be a mixture of a first-day warm-up transient and a
steady-state regime. All statistics are observational, are reported
under an explicit independence caveat that Section VII.B quantifies, and
do not claim identification of any causal or counterfactual quantity.

\section{XIII. Data Availability}\label{xiii.-data-availability}

The companion dataset is deposited on Zenodo under concept DOI
10.5281/zenodo.20043515. The deposit contains rejections.csv (2,997
rows), rejection\_outcomes.csv (67,000 rows), a checksums.txt file whose
SHA-256 values match the files bit for bit, and a dataset README that
describes the schema, the observation environment, the anonymisation of
filter and source labels, and the window boundaries.

\section{XIV. Code Availability}\label{xiv.-code-availability}

The reference implementation (prfs v2.0.0) and the secondary cross-check
aggregator (oracle/oracle\_aggregate.py) are deposited alongside the
dataset in the reproducibility package. The package contains the pytest
suite (including tests/test\_reason\_ledger.py, tests/test\_restart.py,
and tests/test\_oracle\_agreement.py), an independent stdlib recompute
script (scratchpad/p2\_repro/independent\_repro.py) that verifies every
headline number without importing prfs or oracle, and a CI configuration
that reruns all tests on the frozen CSVs. Executing python -m prfs.cli
-{}-registry-csv data/rejections.csv -{}-outcomes-csv
data/rejection\_outcomes.csv -{}-report-out reports/verify\_report.json
regenerates every headline number in this manuscript.

\section{XV. Author Contributions, Competing Interests, and
Funding}\label{xv.-author-contributions-competing-interests-and-funding}

\textbf{Author contributions.} Arati Uday Kamat is the sole author of
this paper and is responsible for the methodology, the reference
implementation, the cross-check aggregator, the data deposit, and the
manuscript.

\textbf{Competing interests.} The author declares no competing financial
or non-financial interests related to this work.

\textbf{Funding.} This work received no external funding. All
computation, data collection, and manuscript preparation were
self-funded.

\section{XVI. References}\label{xvi.-references}

\begingroup
\sloppy
\setlength{\parindent}{0pt}
\setlength{\parskip}{4pt plus 1pt minus 1pt}
\begin{hangparas}{2em}{1}
Bailey, David H., and Marcos Lopez de Prado. 2014. ``The Deflated Sharpe
Ratio: Correcting for Selection Bias, Backtest Overfitting, and
Non-Normality.'' SSRN Working Paper 2460551.
https://doi.org/10.2139/ssrn.2460551.

Coston, Amanda, Alan Mishler, Edward H. Kennedy, and Alexandra
Chouldechova. 2020. ``Counterfactual Risk Assessments, Evaluation, and
Fairness.'' In Proceedings of the 2020 Conference on Fairness,
Accountability, and Transparency, 582-593.
https://doi.org/10.1145/3351095.3372851.

Crook, Jonathan, and John Banasik. 2004. ``Does Reject Inference Really
Improve the Performance of Application Scoring Models?'' Journal of
Banking and Finance 28 (4): 857-874.
https://doi.org/10.1016/j.jbankfin.2003.10.010.

Daian, Philip, Steven Goldfeder, Tyler Kell, Yunqi Li, Xueyuan Zhao,
Iddo Bentov, Lorenz Breidenbach, and Ari Juels. 2020. ``Flash Boys 2.0:
Frontrunning in Decentralized Exchanges, Miner Extractable Value, and
Consensus Instability.'' In 2020 IEEE Symposium on Security and Privacy,
910-927. https://doi.org/10.1109/sp40000.2020.00040.

Fang, Fan, Carmine Ventre, Michail Basios, Leslie Kanthan, David
Martinez-Rego, Fan Wu, and Lingbo Li. 2022. ``Cryptocurrency Trading: A
Comprehensive Survey.'' Financial Innovation 8 (1): 13.
https://doi.org/10.1186/s40854-021-00321-6.

Kamat, Arati. 2026a. ``Outcome-Classified Precision Auditing of Filter
Rules in Algorithmic DEX Trading: Evidence from 2,400 Rejection
Events.'' SSRN Working Paper 6638259.

Kamat, Arati. 2026b. ``RED-2400: A Public Benchmark of
Algorithmically-Rejected Trading Events with Outcome Labels.'' SSRN
Working Paper 6702198.

Lakkaraju, Himabindu, Jon Kleinberg, Jure Leskovec, Jens Ludwig, and
Sendhil Mullainathan. 2017. ``The Selective Labels Problem: Evaluating
Algorithmic Predictions in the Presence of Unobservables.'' In
Proceedings of the 23rd ACM SIGKDD International Conference on Knowledge
Discovery and Data Mining, 275-284.
https://doi.org/10.1145/3097983.3098066.

Mancino, Davide. 2025. ``The Memecoin Phenomenon: An In-Depth Study of
Solana's Blockchain Trends.'' In 2025 IEEE Symposium on Computers and
Communications (ISCC). https://doi.org/10.1109/iscc65549.2025.11326251.

Wilson, Edwin B. 1927. ``Probable Inference, the Law of Succession, and
Statistical Inference.'' Journal of the American Statistical Association
22 (158): 209-212. https://doi.org/10.1080/01621459.1927.10502953.

\begin{center}\rule{0.5\linewidth}{0.5pt}\end{center}

\section{Figures}\label{figures}

\begingroup
\par\noindent\centering
\includegraphics[width=0.9\linewidth]{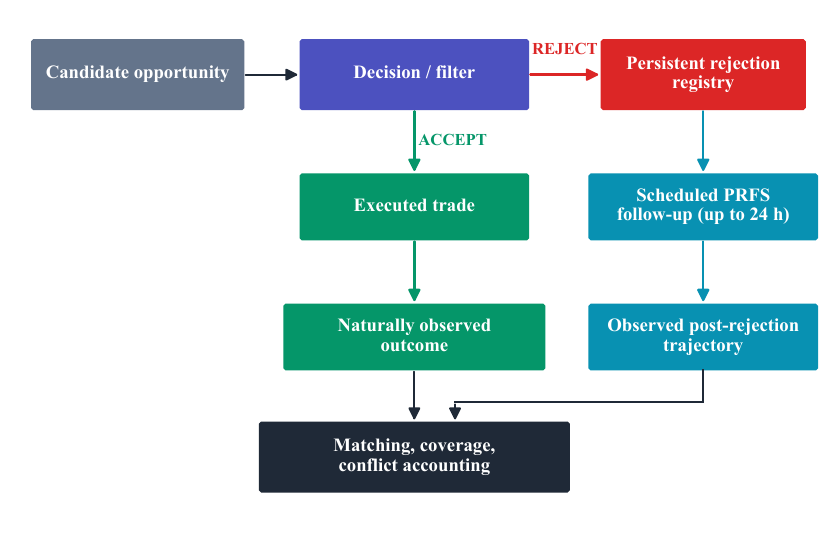}
\par
\endgroup

\textbf{Figure 1. PRFS system architecture and event flow.} Candidate
opportunities are evaluated by the filter-gated trading system. Accepted
candidates proceed through the execution path, where outcomes are
naturally observed. Rejected candidates are recorded in the persistent
rejection registry and scheduled for PRFS follow-up sampling. The
resulting post-rejection trajectories are subsequently matched to
rejection events for coverage and reason-conflict accounting.

\textbf{Figure 2.} Per-event forward-sample-density histogram. For each
distinct outcome three-field key (mint, rejectTs, rejectReason) the
number of scheduled sample rows is counted; the grey histogram covers
all distinct outcome three-field keys, and the blue overlay covers the
subset that match a registry event on the primary three-field key. The
grey excess above the blue bars corresponds to the orphan-key population
documented in Section VIII.H. Rendered from
figures/fig2\_sample\_density.pdf and regenerated deterministically from
the frozen deposit by figures\_gen.py.

\textbf{Figure 3.} Per-filter registry event volume and matched-event
overlay. Grey bars: registry events per filter label (sums to 2,997).
Blue bars: registry events per filter label that match at least one
outcome row on the three-field key (sums to 1,455). Percentage labels
above each blue bar are the per-filter coverage rates from Table 1.
Rendered from figures/fig3\_filter\_volume.pdf and regenerated
deterministically from the frozen deposit by figures\_gen.py.
\end{hangparas}
\endgroup

\end{document}